\documentstyle[preprint,aps,eqsecnum]{revtex}

\def\beq#1{\begin{equation} \label{#1}}
\def\eeq{\end{equation}}

\def\bra#1{\left\langle #1\right\vert}
\def\ket#1{\left\vert #1\right\rangle}
\def\epsp{\epsilon^{\prime}}
\def\NPB{{ Nucl. Phys.} B}
\def\PLB{{ Phys. Lett.} B}
\def\PRL{ Phys. Rev. Lett.}
\def\PRD{{ Phys. Rev.} D}

\begin{document}
{
\tighten

\title{Why do neutrinos with different masses interfere and oscillate?
Why are states with different masses but
same energy coherent?
Overcoming barrier between particle \& condensed matter physics}

\author{Harry J. Lipkin\,\thanks{Supported in part by
U.S.
Department of Energy, Office of Nuclear Physics, under contract
number
DE-AC02-06CH11357.}}
\address{ \vbox{\vskip 0.truecm}
  Department of Particle Physics
  Weizmann Institute of Science, Rehovot 76100, Israel \\
\vbox{\vskip 0.truecm}
Physics Division, Argonne National Laboratory
Argonne, IL 60439-4815, USA\\
~\\harry.lipkin@weizmann.ac.il
\\~\\}

\maketitle

\begin{abstract}
Neutrino oscillations occur only if it is impossible  to determine $\nu$ mass by using conservation laws on measurements of nucleon-lepton system absorbing  $\nu$. No oscillations if $\nu$ detector is mass spectrometer. Beam is split into components with different
masses entering different counters. For each event only one counter will click and determine $\nu$ mass. Condensed matter physics needed to describe the $\nu$ detector, show it is not a mass spectrometer and identify which properties of the incident $\nu$ are unobservable.
Relativistic quantum field theory can only describe $\nu$ wave function entering detector but not large  uncertain momentum transfers to detector nor associated energy-momentum asymmetry.
Absorption of incident $\nu$'s with different momenta but same energy leaves no trace of initial $\nu$ momentum difference in finite-size $\nu$ detector with effectively infinite mass at rest in laboratory. Undetectable recoil-free momentum is transferred to the detector with negligible energy transfer. The Debye-Waller factor common in X-ray diffraction by crystals gives probability that absorbing $\nu$'s with different momenta produce same nucleon-charged-lepton final state.
Oscillations in time described in  textbooks as interference between $\nu$ states with different
energies not observable in realistic experiments.  Different energy $\nu$'s not coherent because energy can be determined by measurements on initial and final states.
Experiments detecting $\nu$ produced by
$\pi \rightarrow \mu \nu$ decay observe no electrons even though $\nu$ mass eigenstates produce electrons.
Electron amplitude canceled by interference between  amplitudes from different $\nu$ mass eigenstates with same energy and different momenta entering massive detector.
\end{abstract}
}

\def\beq#1{\begin{equation} \label{#1}}
\def\eeq{\end{equation}}
\def\bra#1{\left\langle #1\right\vert}
\def\ket#1{\left\vert #1\right\rangle}
 \def\epsp{\epsilon^{\prime}}
\def\NPB{{ Nucl. Phys.} B}
\def\PLB{{ Phys. Lett.} B}

\def\PRL{ Phys. Rev. Lett.}
\def\PRD{{ Phys. Rev.} D}

\section {
Quantum mechanics missed at
particle - condensed matter interface}

\subsection{A communications gap  that deserves serious thought}

Consider how a linear combination of $\nu$ waves having different masses are absorbed by a nucleon in a detector.
Standard particle physics treatments describe only the coherence properties of the wave function entering the detector.
But this is only part of the story. Understanding what happens after the wave function enters the detector is crucial
to explain why neutrino oscillations are observed.

Suppose that the neutrino detector is a neutrino mass spectrometer. The beam will be split into components with different
masses entering different counters. For each event only one counter will click and determine the $\nu$ mass. There can
be no observable oscillations if the detector is a mass spectrometer.

That neutrino oscillations are observed tells us the $\nu$ detector is not a mass spectrometer. Then what is it?
What are the properties of the detector that prevent the determination of the mass of the $\nu$ that entered the
detector?

Here we need condensed matter physics. The detectors in all neutrino oscillation experiments are condensed matter systems with effectively infinite mass at rest in the laboratory and in thermal equilibrium with its environment. The neutrino is absorbed by a nucleon with the emission of a charged lepton. The nucleon
bound in this infinite mass system can absorb two incident neutrino waves with different momenta and produce
transitions to the same final state of the detector nucleon. The two momenta are absorbed with the same energy transfer.
The momentum difference is absorbed by the infinite mass detector leaving no trace of the momentum difference.
The probability of absorbing two waves with different momenta in transitions to the same final state depends
upon the relative phase of the two waves. Studying transitions as a function of the position of the detector is
studying them as a function of the relative phase. Oscillations are observed when two neutrino waves with different
momenta and the same energy are detected at different distances from the neutrino source.
This immediately tells us that interference is observable only between neutrino states with the same energy and
different momenta. States with different energies will not interfere.

This can be summed up as follows:

\begin{enumerate}
          \item 
          No coherence between the amplitudes  having different masses if any measurement on the final state of the detector can determine the neutrino mass.
          \item When a two neutrino waves with slightly different momenta are  absorbed by a nucleon bound in a large system with effectively infinite mass, there is a probability that the same bound nucleon final state is produced with the momentum difference absorbed by the large system. The probability that this can occur can be calculated by knowing the wave function of the absorbing nucleon. This probability is called
 the "Debye-Waller" factor in condensed matter physics.
             \item
             Interference between absorption of two incident waves with same energy and different momenta leave no observable trace of momentum difference in final state.
           \end{enumerate}

            This simple physics encounters a communications barrier because:
 \begin{itemize}
 \item Particle physicists are sure that neutrino oscillations can be described  by single-particle quantum  field theory and  it is not necessary to know any condensed matter physics, They have probably
 never heard of a Debye-Waller factor and unaware that the presence of interactions with a large system has any relevance to particle physics. They never discuss why neutrino oscillations are observable in real experiments.
 \item Condensed matter physicists show no interest in $\nu$ oscillations. This is useless particle physics.
\end{itemize}
\subsection{Nucleon binding prevents determination of momentum transfer}

Neutrino oscillations cannot be produced by interference between two $\nu$ states with different masses if measurements on the nucleon-lepton system absorbing the $\nu$ can determine the $\nu$ mass. The nucleon is bound in a quantum-mechanical condensed matter system with well defined energy levels.  Neutrino oscillations are observed because the
nucleon final state is not a momentum state and the momentum spread prevents determination of the $\nu$ mass. An oscillation
wave length cannot be measured unless the size of the detector is much smaller than the wave length of the
oscillation to be measured. This localization of the detector nucleon produces a momentum spread which prevents the use of momentum conservation in determining the neutrino mass. There is no energy spread because the transition occurs between discrete energy levels. This momentum spread without energy spread occurs only in the laboratory system.

This uncertainty only in momentum and not in energy in the nucleon-lepton system is not relativistically covariant. Lorentz transformations are useless here because they boost the whole detector. Relativistic quantum field theory can provide a full description of the neutrino wave function incident on  a detector. But it cannot describe which coherence properties of the wave function in the laboratory system remain in the nucleon-lepton system after $\nu$ absorption in the detector.
\subsection{The $\nu$ absorption by the detector is described by condensed matter physics}

The Debye-Waller factor common in X-ray diffraction will be shown below to give the probability that two components of an incident neutrino with a momentum difference $\delta p$ can produce the same final quantum state in the detector.
\beq{DW}
f_{DW}=e^{- (\delta p)^2 \langle X^2\rangle}
\eeq
where $\langle X^2\rangle$ denotes the mean square distance of detector nucleon position in the detector from its equilibrium position. The transition leaves no trace of the momentum difference between components of the initial neutrino with different masses.

The physics of neutrino absorption and X-ray diffraction seem to be very different. But the physics  is the same for energy and momentum transfer when a single neutrino or photon is scattered by a massive object. Energy is conserved; momentum is not. The interference fringes in X-ray diffraction and neutrino oscillations both arise because no recoil momentum can identify the position of the momentum transfer. These allow two components of the initial photon or neutrino wave functions with the same energy and different momenta or different paths in configuration space to interfere coherently in producing the observed final state.

In an experiment where $\nu$ oscillations are observed the size of the detector denoted by a length $L$ must be much smaller than the oscillation wave length.
\beq{constraint}
\delta p^2\cdot L^2 \ll 1; ~ ~ ~f_{DW}\gg e^{- (\delta p)^2 L^2}
\eeq
\subsection{How to solve coherence problems in quantum mechanics}

The main question in understanding neutrino oscillations is what prevents the oscillation experiments from being missing mass experiments. The answers are wave-particle duality and coherence.

For a correct description of quantum coherence, consider a two-slit experiment in which an electron passes through a barrier with two slits. The electron is a wave which passes through the screen. the waves coming from the two slits combine and interfere. The interference pattern observed on a screen shows that both parts of an electron wave passed through the slits. The point where the electron will be observed as a particle on a screen cannot be predicted. Only the probability that a single electron will be observed on
the screen can be predicted. But this interference requires that no information can be available to determine through which slit the electron passed.

An essential feature of this coherence is wave-particle duality. The passage of the electron through the slits is described as propagation of the electron as a wave. But the detection on the screen requires the appearance of the electron as a particle. Quantum mechanics describes the electron by a wave function that satisfies the Schroedinger equation and expresses both the wave and particle properties.

Neutrino oscillations are created in a two-slit experiment in in momentum space. The neutrino is described by a wave function which contains components with different masses and different momenta. It is detected by a nucleon in a massive detector and emerges as a lepton with definite energy and momentum. We now investigate how two components of the neutrino wave with different momenta can interfere and produce a nucleon and lepton pair with a single momentum.
A correct description of this process is obtained by writing down the Schroedinger equation and solving it. But in neutrino detection by a massive detector the Schroedinger equation is so complicated that it is not easily solved. Condensed matter physicists have treated this question many times by finding a small parameter and writing the solution as a power series in the small parameter. In most cases the leading term in the series is sufficient for most purposes.
In neutrino oscillations the obvious small parameter is the ratio of the detector size to the oscillation
wave length. We sho w below  that the leading term in an expansion in this small parameter is adequate.

The application of quantum mechanics to problems of coherence and interference has been confused by the
Schroedinger cat paradox. There the cat is alive in a cage and is killed by the arrival of a bullet shot from a gun triggered by the decay of a radioactive nucleus. The suggestion that the cat is in a coherent mixture of life and death is simply wrong. The cat is either alive or dead even if no one opens the cage. The radio active nucleus is not an isolated quantum system. It is coupled to an environment which will change drastically when the nucleus decays, changes its state and emits an $\alpha$, $\beta$ or $\gamma$ ray. One does not need to open the cage to see if the cat is alive or dead. The environment always contains the information.

We now explore in detail the implications of an unobserved recoil momenta using a
model-independent description of the detector nucleon absorbing the neutrino. It applies not only for
the case where the detector nucleon is bound in a crystal\cite{LipQM,CIPANP,pedneut,Dolgov} but also for amorphous solid,
liquid and gas detectors where the detector is confined to a region of linear dimension $L$ satisfying
(\ref{constraint}).

\section {Detailed analysis of
effects of observed momentum transfer}

\subsection {Coherence in absorption of two $\nu$ states with different masses}

Consider the amplitudes for absorption of two neutrino states with slightly different
masses.  If energy and momentum are exactly conserved in the nucleon-lepton system, the two final nucleon-lepton states
are orthogonal. There is then no interference and there are no oscillations. There is interference only if no measurement
on the final state can determine the momentum of the entering neutrino. To have interference there needs to be a
momentum transfer to the whole detector which is just enough to cover up the change denoted by $\delta \vec p$ in
the momentum of the entering neutrino.

The detector is a large many-body system at rest in the laboratory with a discrete energy level spectrum.     Consider the transition matrix element $\bra{B}T(\vec p)\ket{A}$ for the case where the detector makes a transition
between an initial detector eigenstate denoted by $\ket{A}$ to a final state denoted by $\ket{B}$ which differs from the initial state by the absorption of a $\nu$ with momentum $\vec p$, a nucleon charge change and
 the emission of a charged lepton.
The transition between the same initial detector eigenstate $\ket{A}$ to the same final state $\ket{B}$ by the absorption of a $\nu$ with momentum $\vec p+ \delta \vec p$ is denoted by
\beq{abtrans}
\bra{B}T(\vec p + \delta \vec p)\ket{A} =
\bra{B}T(\vec p)e^{i\delta \vec p\cdot \vec X}\ket{A} \approx
\bra{B}T(\vec p)\cdot [1-
\frac{1}{2}\cdot[\delta \vec p^2\cdot \vec X^2]
\ket{A}
\eeq
where $\vec X$ denotes the distance between the position of the nucleon
and the center of the detector.
For a small change $\delta \vec p$
\beq{deltrans}
\frac{\bra{B}T(\vec p + \delta \vec p)\ket{A}-
\bra{B}T(\vec p)\ket{A}}{\bra{B}T(\vec p)\ket{A}}
\approx -\frac{1}{2}\cdot
\frac {\bra{B}[\delta \vec p^2\cdot \vec X^2]\ket{A}}{\bra{B}T(\vec p)\ket{A}}
\leq \frac{\delta p^2\cdot L^2}{2}
\eeq
where we have taken the leading term in the expansion of the small parameter $\delta p$. and $L$ denotes the length of
the detector.
 We see that a change by an amount $\delta p$ in the neutrino momentum will not be detected by measuring the
transition $\ket{A}\rightarrow \ket{B}$  as long as
the size of the detector satisfies (\ref{constraint}) and is much smaller than the oscillation wave length.
If absorption of two neutrino states with slightly different momenta can produce the same
change from energy level A to energy level B there is coherence. One only sees that there was
a transition from A to B.
One cannot know which neutrino mass produced the transition.
The momentum difference is taken up by the whole detector.

The derivation of (\ref{deltrans}) is independent of the model for the detector. It applies not only for
the case where the detector nucleon is bound in a crystal\cite{LipQM,CIPANP,Dolgov} but also for amorphous solid,
liquid and gas detectors where the detector is confined to a region of linear dimension $L$ satisfying
(\ref{constraint}).

\subsection {Application to detection of $\nu$'s
produced in $\pi \rightarrow \mu\nu$ decay}
We now apply  eqs.(\ref{abtrans} - \ref{deltrans}) to treat and explain the observation that no electrons are
produced in the detector.
We consider two neutrino mass eigenstates, denoted by $\nu_1$ and $\nu_2$ with momenta
$\vec p_\nu$ and $(\vec p_\nu+\delta \vec p)$ and include the $\nu \rightarrow e$ transition.

\beq{abtransb}
\bra{B(\vec p_A+\vec p);e(\vec p_e)}T\ket{A(\vec p_A);\nu_1(\vec p_\nu)} =
\bra{B}e^{i\vec p\cdot \vec X}\ket{A}\cdot \bra{e(\vec p_e)}T_W\ket{\nu_1(\vec p_\nu)}
\eeq

\beq{abtransc}
\bra{B([\vec p_A+\vec p+\delta \vec p]);e(\vec p_e)}T\ket{A(\vec p_A);\nu_2(\vec p_\nu)+\delta \vec p} =
\bra{B}e^{i[\vec p+\delta \vec p]\cdot \vec X}\ket{A}\cdot \bra{e(\vec p_e)}T_W\ket{\nu_2([\vec p_\nu + \delta \vec p])}
\eeq
where $ \vec p_A, \vec p_e$ and $\vec p_\nu$ denote the momenta of the initial nucleon state, the final electron and
the incident neutrino and $T_W$ is the interaction producing the weak transition, Using eq (\ref{deltrans})
and neglecting the small parameter $\delta p^2\cdot L^2$ gives
\beq{abtransf}
\frac{\bra{B([\vec p_A+\vec p+\delta \vec p]);e(\vec p_e)}T\ket{A(\vec p_A);\nu_2(\vec p_\nu +\delta \vec p)}}
{\bra{B(\vec p_A+\vec p);e(\vec p_e)}T\ket{A(\vec p_A);\nu_1(\vec p_\nu)}}\approx
\frac {\bra{e(\vec p_e)}T_W\ket{\nu_2(\vec p_\nu+\delta \vec p}}{\bra{e(\vec p_e)}T_W\ket{\nu_1(\vec p_\nu)}}\approx
\frac {\bra{\nu_e}\nu_2\rangle}{\bra{\nu_e}\nu_1\rangle}
\eeq
Where we note that the ratio of the weak transition matrix elements is equal to the ratio of the elements of the
flavor-mass mixing matrix denoted by $\bra{\nu_e}\nu_2\rangle$ and $\bra{\nu_e}\nu_1\rangle$ and neglect the
dependence of the weak transition matrix element on the small momentum difference $\delta p$.

Consider an incident neutrino $\nu_i$ which is a linear combination of the two mass eigenstatates
\beq{nui}
\ket{\nu_i} = \sum_{k=1}^2  \ket{\nu_k} {\bra{\nu_k}\nu_i\rangle}
\eeq
\beq{abtransg}
\frac{\bra{B([\vec p_A+\vec p+\delta \vec p]);e(\vec p_e)}T\ket{A(\vec p_A);\nu_i(\vec p_\nu+\delta \vec p)}}
{\bra{B(\vec p_A+\vec p);e(\vec p_e)}T\ket{A(\vec p_A);\nu_1(\vec p_\nu)}}\approx
\sum_{k=1}^2
\frac {\bra{\nu_e}\nu_k\rangle\cdot \bra{\nu_k}\nu_i\rangle}{\bra{\nu_e}\nu_1\rangle}=
\frac {\bra{\nu_e}\nu_i\rangle}{\bra{\nu_e}\nu_1\rangle}
\eeq

The probability that an incident neutrino $\nu_i$ is absorbed with electron emission is seen to vanish
if $\nu_i$ is just the right mixture of mass eigenstates to be orthogonal to the electron neutrino state $\nu_e$.
This explains the failure to observe electrons in the detection of neutrinos from $\pi \rightarrow \mu \nu$ decays.

\section {Oscillations can arise only if $\nu$ mass is unobservable}
 \subsection{Momentum and energy conservation violations in the lepton-nucleon system}

 The wave function of a $\nu$ emitted in a weak decay is a linear combination
of states containing different $\nu$ masses, energies and momenta. The $\nu$ is  observed
in a detector by an interaction which changes the charge of a nucleon and emits a charged lepton.
\beq{chex}
\nu + P \rightarrow \mu^+ + N; ~ ~ ~ \nu + N \rightarrow \mu^- + P;  ~ ~ ~
\nu + P \rightarrow e^+ + N; ~ ~ ~ \nu + N \rightarrow e^- + P
\eeq
 $\nu$ oscillations
and the failure to observe electrons in $\pi \rightarrow \mu\nu$ decay\cite{leder} can occur only
if there is coherence and interference between components of the neutrino wave function with
different neutrino masses.
If oscillations are observed something must prevent knowing the neutrino mass
The neutrino mass is measurable if the lepton energy and the energy and momentum changes in the detector nucleon are all observable. They are all observable if:

\begin{enumerate}
\item The neutrino energy is equal to the sum of the change in detector nucleon energy and the lepton energy
\item The neutrino momentum is equal to the sum of the change in detector nucleon momentum plus the lepton momentum
\end{enumerate}

The $\nu$ mass is determined in this ``missing mass" experiment.
If the $\nu$ mass can be determined by measurements on the initial and final states there can be no interference and no oscillations.

But neutrino oscillations are observed. What is wrong with this argument?

To observe oscillations the position of the detector must be known with an error much less that the oscillation wave length. Heisenberg position-momentum uncertainty prevents the
neutrino momentum from being known with sufficient precision to determine the neutrino mass.
The interaction between the neutrino and a detector with effectively infinite mass allows a finite unobserved momentum to be transferred to the detector without energy transfer. This is the same physics as the recoilless momentum transfer in photon scattering by a crystal in X-ray crystallography. Thus:
\begin{enumerate}
\item The neutrino energy IS equal to the sum of the change in detector nucleon energy and the lepton energy
\item The neutrino momentum IS NOT equal to the sum of the change in detector nucleon momentum plus the lepton momentum
\begin{itemize}
\item The nucleon is not free but bound in a large system with effectively infinite mass.
\item The system can absorb momentum without energy transfer.
\item This ``missing momentum" prevents the determination of the neutrino mass
\end{itemize}
\item Neutrino absorption is not a ``missing mass" experiment.
\end{enumerate}
 \subsection {Text books are misleading. No connection with real experiments}

Neither text books nor relativistic quantum field theory tell us what is unobservable.
 Some energy or momentum must be unobservable to produce oscillations.

 Text books tell us that a $\nu$ at rest  with definite flavor is a coherent mixture of
energy eigenstates. Interference between these states produces oscillations
in time between different flavors. The $\nu$'s oscillate as coherent mixtures
of states with different energies.
Text books don't tell us how such an oscillation is created
or observed in any real experiment.

\begin{enumerate}
\item No experiment has ever seen a $\nu$ at rest
\begin{itemize}
\item Neutrinos observed in experiments have no unique rest frame
\item Components of $\nu$ wave function with different masses have different rest frames.
\end{itemize}
\item $\nu$ 's with different energies cannot interfere because their energy is measurable.
\item Detectors in experiments observing $\nu$ oscillations do not measure time
and destroy all interference between states with different energies
\end {enumerate}

What is observable depends on the quantum mechanics of the detector.

\subsection{No oscillations in a ``missing mass" experiment}

The original Lederman-Schwartz-Steinberger experiment\cite{leder}  found that the neutrinos
emitted in a $\pi-\mu$ decay produced only muons and no electrons. Experiments
now show that at least two neutrino mass eigenstates are emitted in $\pi-\mu$
decay and that at least one of them can produce an  electron in a neutrino
detector. The experimentally observed absence of electrons can be explained
only if the electron amplitudes received at the detector from different
neutrino  mass eigenstates are coherent and exactly cancel.

The neutrinos are linear combinations of mass
eigenstates with different masses, different energies and different momenta.
The detector must know that the relative phases of relevant amplitudes will
cancel the production of an electron.
This can only be understood by investigating the quantum mechanics of the detector.

A missing mass experiment was not  performed.

 \subsection {X-Ray and M\"ossbauer physics needed to understand interference in $\nu$ detectrion}
\begin{enumerate}
\item The detector has a definite position in the laboratory system for all times and has effectively infinite mass.
\item Energy in the laboratory system is conserved; momentum conservation is violated
as in X-ray diffraction by crystals and the M\"ossbauer effect\cite{LipQM,CIPANP}.
\begin{itemize}
\item In the M\"ossbauer effect a photon is scattered by an atom in a crystal.
Energy in the laboratory system is conserved. A
missing recoil momentum is absorbed by the crystal with negligible energy loss.
\item In $\nu$ experiments the $\nu$ is absorbed by a nucleon in a detector.
Energy in the laboratory system is conserved.
A missing recoil momentum is absorbed by the detector with negligible energy loss.
\item The same quantum state of the crystal or detector is produced by transitions with different momentum transfers and the same energy transfer.
\item No measurement on the final state can determine momentum of the initial photon or $\nu$.
\end {itemize}
\item The detector can absorb $\nu$'s with a small momentum difference and the same energy transfer and produce the same final state.
\item Momentum difference and mass difference between two $\nu$ states is not observable.
\end{enumerate}

\section {Basic quantum mechanics of  coherence in $\nu$ detection}

\subsection{Quantum mechanics of the detector}

In neutrino oscillation experiments the neutrino is absorbed by a detector which is a complex interacting many-body system at rest in the laboratory. It is described in quantum mechanics by a Hamiltonian. Diagonalizing this Hamiltonian gives a discrete energy spectrum.
Two neutrinos with different masses incident on the detector can be absorbed coherently only if they produce exactly the same final state wave function of the many-body system. Since the energy spectrum is discrete, only neutrinos with the same energy can be absorbed coherently and produce oscillations. Coherence and interference arise when neutrino eigenstates  with different masses and momenta but the same energy are absorbed by a detector and produce the same transition from a given set of detector energy eigenstates. How this occurs is treated explicitly below.

\subsection{Difference between space and time measurements}

In neutrino experiments the distance between source and detector can be measured with arbitrary precision. The transit time between source and detector cannot. The neutrino arrives at the detector as a finite wave packet over an appreciable time interval. The detection can occur at any time that a portion of the wave is still in the detector. This gives an uncertainty in transit time which cannot be shortened. Furthermore, if the neutrino wave packet contains components with different masses, the centers of the two wave packets arrive at the detector at different times. This time difference has produced a controversy about a factor of two.

The proper way to treat this problem is to note that oscillations can occur only if the time intervals for the two wave packets still overlap almost completely and interference can occur. The time of an individual neutrino arrival can be measured. The probability for emission of a charged lepton with a definite flavor depends on the relative magnitudes and phases of the contributing mass eigenstates at this time. The length of the wave packet in time is sufficiently short so that the relative phases and the flavor of the emitted charged lepton  do not change appreciably with the exact time of observation. However the relative phase does change with changes on the position of the detector. This phase change with distance produces neutrino oscillations.

It is well known that at long times like the times of arrival from a supernova the wave packets separate and there is no interference and no oscillations. We consider here intermediate times which are long enough to produce oscillations and short enough so that there is nearly complete overlap between the wave packets of neutrinos with different eigenstates and different velocities.

This asymmetry between time and distance is essential for understanding neutrino oscillations and is not easily treated by relativistic quantum field theory.
\subsection {Constraints on the detector nucleon wave function}
Neutrino oscillations can be observed only if the detecting nucleon is confined for all
times to a region of space in the laboratory system much smaller than the oscillation wave length.

The probability of finding the detector nucleon outside the detector must vanish
for all times. The state of the detector nucleon in quantum mechanics is described by a wave function
or density matrix which gives a time-independent vanishing probability for finding
the nucleon outside the detector. The density matrix describing the detector nucleon
must have coherence and interference between components with different momenta
at each energy which cancel out the probability of finding the nucleon outside
the detector.
\subsection {Implications of space-time condition on the detector}

This space-time condition on the detector nucleon wave function
is crucial for a description of $\nu$ oscillations, missed in theoretical investigations, e.g.
\cite{Ahmedov,Zoltan}.
and not included in formulations based on quantum field theory. Only components of the
incident $\nu$ wave function with the same energy and different momenta are coherently absorbed,
produce the same transitions between two detector nucleon
eigenstates and interfere to create the observed oscillations.
Since the $\nu$ momenta producing these transitions are not observable
$\nu$ absorption is a which-path experiment
in momentum space. Oscillations in configuration space are produced by
interference between final states with same energy and different momenta.
The experimental observation of $\nu$ oscillations shows that the
$\nu$ wave function entering the detector
contains coherent linear combinations of states with same energy, different
momenta and definite relative phases\cite{gsihjl}.

\begin{enumerate}

\item Components of an incident neutrino with the same energy and different
momenta can produce coherent transition amplitudes between two detector nucleon
states that both have a vanishing probability of finding the nucleon outside
the detector.

\item The momentum of the neutrino that produced the transition in the detector
is not observable.

\item Neutrino detection is a ``two-slit" or ``which-path" experiment in momentum
space.

\end {enumerate}
\subsection {A ``which path" experiment with Dicke superradiance}

Absorption of two $\nu$ states with different momenta gives the same detector transition.
Dicke superradiance\cite{Super} arises when several initial states produce same final state.
The transition matrix element depends upon the relative phases of these amplitudes.
The state with maximum constructive interference is called ``superradiant".
The states orthogonal to the superradiant state are called ``subradiant"
The relative phase between states having different momenta changes with the distance
between source and detector.
This phase change gives transitions between superradiant and subradiant states and
produces the observed oscillations.

Neutrino detection is a ``which-path" or ``two-slit" experiment in momentum space.
Dicke superradiance explains oscillations with distance.
Momentum conservation is violated in the nucleon-lepton system by $\nu$ absorption on
a nucleon in the detector.
The detector absorbs the missing momentum - like crystal in the M\"ossbauer effect.
Absorption of two $\nu$ states with different momenta produce the same detector transition.

That the error in the neutrino momentum is sufficiently large to allow neutrino oscillations
is easily seen. The size of the detector must be much smaller than the wave
length of the oscillation in space in order for oscillations to be observable. This implies that
the difference in momenta of the interfering neutrino waves is much smaller that the spread in the
momentum of the detector.
\subsection {Energy-momentum asymmetry crucial to understanding $\nu$ oscillations}
An energy-momentum asymmetry not treated in covariant treatments
arises from the asymmetry between space and time in the detector nucleon wave function.
The probability for finding nucleon outside detector spatial region vanishes for all times.
The detector nucleon wave function must then vanish in space outside detector for all times.
\begin{itemize}
\item Components of the wave function at each energy must cancel outside the detector
\item Components with the same energy and different momenta can be coherent
\item Interference between states with different energies and same momentum cannot vanish outside detector
\item Absorption of $\nu$'s with different momentum and same energy can be coherent.
\end{itemize}

This crucial energy-momentum constraint is valid only in the laboratory frame.
Covariant treatments and relativistic quantum field theory cannot explain this constraint.


\section{Why different approaches give the same answer}

Consider
a simplified two-component $\nu$ state with two
components having momenta $\vec P$ and $\vec P +\delta \vec P$ with
energies $E$ and $E +\delta E$ and squared masses $m^2$ and $m^2+ \Delta (m^2)$.
Changes produced by a small change $\Delta (m^2)$ in the squared neutrino
mass satisfy the relation
\beq{epconsdel}
E^2 = P^2 + m^2; ~ ~ ~ ~ 2E\delta E \approx  2P\delta P + \Delta (m^2)
\eeq
When these neutrinos travel through a distance $X$ in a time $t$ their phase is
given by:
\beq{phases}
\phi(E,P) = P\cdot X -E\cdot t; ~ ~ ~\phi(E+\delta E,P+ \delta P) =
(P+ \delta P)\cdot X -  E(E+\delta E)\cdot t
\eeq

Realistic experiments detect the $\nu$ at a known and definite distance $X$ from the source.
The time $t$
of detection of the   $\nu$ is not measured. The
transit time $t_w$ of the center of the
wave packet traversing the distance $X$ can be  estimated using the
group velocity $v_g$ of the  neutrino wave packet

\beq{gruvel}
 v_g = \frac {P}{E}; ~ ~ ~  t_W= \frac {X}{v_g}  = \frac {E}{P}\cdot X
\eeq

The relative phase $\delta \phi(X)$ between the two components observed at point $X$ is
\beq{phasediff}
\delta \phi(X) = \delta P \cdot X - \delta E\cdot t \approx
X \cdot \left[ \delta P - \delta E\cdot \frac {E}{P}  \right]- \delta E\cdot (t-t_W)\approx
-\frac {X}{2P}\cdot \Delta (m^2)+ 2\delta E\cdot (t-t_W)
\eeq

If $\delta E\cdot (t-t_W) =0$
the relative phase $\delta \phi(X)$ depends only
on the squared mass difference $\Delta (m^2)$ between mass eigenstates.
It is independent of the particular chosen values of $\delta P$  and $\delta E$
as long as these satisfy the relation (\ref{epconsdel})
We note that $\delta E\cdot (t-t_W) =0$ if $\delta E =0$
and we are comparing two components with
the same energy. Treatments where the two have different energies choose $t \approx t_W$
and also have $\delta E\cdot (t-t_W) \approx 0$.
Thus choosing states with the same energy and different momenta gives the same result as
choosing states with the same momentum and different energies.

This explains the apparent miracle that different treatments
of $\nu$ oscillations always give the same answer for the relation between the
oscillation wave length and the mass squared difference between $\nu$ mass
eigenstates.
Quantum fluctuations in the transit time of the wave packet do not affect this
result since neither approach uses the fluctuating value of $t$.

\section{Conclusions}

Neutrino scillations cannot occur if the momenta of all other particles
participating in the reaction are known and momentum and energy are conserved. Heisenberg uncertainty allows
interference between neutrino mass eigenstates with different momentum
transfers out of the nucleon-lepton system. Interference observed in experiments
arises from momentum transfers to a large detector, analogous to the momentum
transfer to a whole crystal in coherent diffraction of X-rays and the M\"ossbauer effect. The wave function of the
nucleon absorbing the neutrino in an oscillation
experiment is confined to a region whose scale is much smaller than oscillation
wave length. The condition that this wave function must vanish for all times
outside this region
in the laboratory system is not properly included in covariant
treatments. It shows that the oscillations are produced by interference between
neutrino states with different masses and momenta but the same energy.
A complete description of the decay process must include the interaction with
the environment and violation of momentum conservation.

\section{Acknowledgement}
The theoretical analysis in this paper was motivated by discussions with  Paul
Kienle at a very early stage of the experiment in trying to understand whether
the effect was real or just an experimental error.
It is a pleasure to thank him for calling my attention to this problem
at the Yukawa Institute for Theoretical Physics at Kyoto  University, where
this work was initiated during the YKIS2006 on ``New  Frontiers on QCD".
Discussions on possible experiments with Fritz Bosch, Walter Henning, Yuri
Litvinov and Andrei Ivanov are also gratefully acknowledged. The author also acknowledges further
discussions on neutrino oscillations as ``which path" experiments with Evgeny Akhmedov,  Eyal
Buks, Avraham Gal, Terry Goldman, Maury Goodman,  Yuval Grossman, Moty Heiblum, Yoseph Imry,
Boris Kayser, Lev Okun, Gilad Perez, Murray Peshkin, David Sprinzak, Ady Stern,
Leo  Stodolsky and Lincoln Wolfenstein.

%
\catcode`\@=11 
\def\references{
\ifpreprintsty \vskip 10ex
%
\hbox to\hsize{\hss \large \refname \hss }\else
\vskip 24pt \hrule width\hsize \relax \vskip 1.6cm \fi \list
{\@biblabel {\arabic {enumiv}}}
{\labelwidth \WidestRefLabelThusFar \labelsep 4pt \leftmargin \labelwidth
\advance \leftmargin \labelsep \ifdim \baselinestretch pt>1 pt
\parsep 4pt\relax \else \parsep 0pt\relax \fi \itemsep \parsep \usecounter
{enumiv}\let \p@enumiv \@empty \def \theenumiv {\arabic {enumiv}}}
\let \newblock \relax \sloppy
 \clubpenalty 4000\widowpenalty 4000 \sfcode `\.=1000\relax \ifpreprintsty
\else \small \fi}
\catcode`\@=12 
{\tighten

\end{document}